\documentclass[12pt]{iopart}

\usepackage{latexsym,graphicx}
\begin{document}

\title[Interlayer coupling in rotationally faulted multilayer graphenes]{Interlayer coupling in rotationally faulted multilayer graphenes}

\author{E. J. Mele}

\address{
Department of Physics and Astronomy \\ David Rittenhouse Laboratory \\ University of Pennsylvania \\ Philadelphia, PA 19104}
\begin{abstract}
This article reviews progress in the theoretical  modelling  of the electronic structure of rotationally faulted multilayer
graphenes. In these systems the crystallographic axes of neighboring layers are misaligned so that the layer stacking does not
occur in the Bernal structure observed in three dimensional graphite and frequently found in exfoliated bilayer graphene.
Notably, rotationally faulted graphenes are commonly found in other forms of multilayer graphene including epitaxial graphenes
thermally grown on ${\rm SiC \, (000 \bar 1)}$, graphenes grown by chemical vapor deposition, folded mechanically exfoliated
graphenes, and graphene flakes deposited on graphite. Rotational faults are experimentally associated with a strong reduction
of the energy scale for coherent single particle interlayer motion. The microscopic basis for this reduction and its
consequences have attracted significant theoretical attention from several groups that are highlighted in this review.
\end{abstract}

\maketitle

\section{Introduction}
Coherent interlayer motion in multilayer graphenes play a crucial role in determining their low energy electronic properties.
In single layer graphene the absence of the layer degree of freedom cleanly exposes the geometrical structure of its low energy
electronic physics. This is controlled by single particle spectra containing linearly dispersing bands around singular points
at its inequivalent zone corners, described by a pair of valley-polarized two dimensional massless Dirac Hamiltonians
\cite{berrygeim,berrykim}. The physics is very different for graphene bilayers that are stacked in the Bernal geometry with the
``A" sublattice of one layer eclipsed with the ``B" sublattice of its neighbor ($AB$ stacking). Here the effects of coherent
interlayer coupling are quite strong and the low energy sector is described instead by a {\it different} class layer-coherent
chiral fermions with a quadratic dispersion and a Berry's phase of $2 \pi$ for reciprocal space orbits that encircle the point
of degeneracy \cite{McCFal}. This physics is readily understood from the experimentally known strength of the interlayer
tunneling amplitude at eclipsed sites and it can be generalized to describe the low energy physics of multi-layer graphenes
where the crystallographic axes of neighboring layers are rotated by special angles $\theta = n \pi/3$
\cite{koshino,minmac,minmac2,pablotrilayer}.

Surprisingly, experimental work over the last five years has revealed a family of multilayer graphenes that show only weak (if
any) effects of their interlayer coupling. This family includes graphenes that are grown epitaxially on the ${\rm SiC \, (000
\bar 1)}$ surface \cite{bergerepitax,deheerreview,haas}, CVD grown graphenes \cite{reina}  and some forms of  exfoliated
graphene \cite{schmidt,STMgraphite,graphenegraphite}. A common structural attribute of these systems is a rotational
misorientation (a twist) of their neighboring layers at angles of $\theta \neq n \pi/3$. The layer decoupling has been inferred
from the measurements of the magnetotransport \cite{bergerepitax,deheerreview}, of the Landau level spectra observed in
scanning tunneling spectroscopy \cite{graphenegraphite,miller} and perhaps most clearly in angle-resolved photoemission spectra
\cite{sprinkle}.

These experimental observations are attracting significant theoretical attention. The layer decoupling in twisted multilayers
is frequently attributed to a kinematical effect whereby the layer projections of the zone corner crystal momenta are {\it
misaligned} by the rotation, preventing momentum-conserving interlayer motion at sufficiently low energy \cite{JMLdS}. In this
scenario the low energy theory is described by four separate valley- and layer- polarized Dirac cones. These fermions are then
re-coupled at a crossover energy scale where the individual Dirac cones merge and hybridize thus changing the band topology
\cite{vanHove}. The simplest version of this theory predicts that the residual low energy effect of the twist is to reduce the
Fermi velocities by an angle dependent factor where the smallest velocities are expected for small rotation angles. These
theoretical predictions have provided a taking off point for further investigations of this problem using a variety of methods
ranging from microscopic atomistic calculations to continuum models designed to capture selected elements of the microscopic
physics. Presently there is a lively discussion concerning the theoretical interpretation of the electronic physics in twisted
graphenes: What is the appropriate long wavelength theory? How can one distinguish between the electronic physics for ``small"
and ``large" rotation angles?  How does the the interlayer coherence scale depend on the fault angle? What are the experimental
consequences of the weak interlayer coherence? It is fair to say that the one-electron physics of these systems is proving to
be unexpectedly rich and it has so far eluded a satisfactory (or at least complete) theoretical description. In this article we
briefly highlight some recent theoretical progress on this problem and focus on some of the major unresolved issues.

Section 2 presents a discussion of the geometric properties of rotationally faulted bilayers which are generally useful for
analyzing their structural and electronic properties. The results presented here provide a foundation for a theoretical
analysis we have presented earlier \cite{ejmrc} though these details have not been published previously. Sections 3-5 briefly
review the existing theoretical approaches that have been developed for describing the electronic structure of these systems.
Section 3 reviews the essential features of a long wavelength theory of layer that illustrates the physics of layer decoupling
by ``rotational mismatch" \cite{JMLdS}.  Section 4 presents some highlights of microscopic atomistic calculations on these
systems and Section 5 briefly reviews the content of several ``second generation" continuum theories that refine the original
theoretical proposal. Section 6 summarizes with a discussion of the connections of these theories to experiment and points to
some interesting open problems.

\section{Geometrical Considerations}

\subsection{Lattice Structures}

A twisted graphene bilayer can be characterized by a relative rotation of the symmetry axes of its two layers through angle
$\theta$ and a rigid translation $\vec \Delta$. Holding one layer fixed, a rotation about the point $\vec r_0$ maps coordinates
$\vec r$ in the fixed layer to positions $\vec r'$ in the rotated layer in the manner
\begin{eqnarray}
\vec  r' = {\cal R} (\theta) \cdot \left(\vec r - \vec r_0 \right) + \vec \Delta
\end{eqnarray}
where ${\cal R}(\theta)$ is the two-dimensional rotation operator. For definiteness one can consider the situation where the
rotation is taken about a lattice site and the relative shift $\Delta=0$. A commensurate rotation occurs when a lattice
translation of the unrotated layer $\vec T_{mn} = m \vec a_1 + n \vec a_2$ spanned by its two primitive layer translations
$\vec a_1$ and $\vec a_2$ and the mapping of an {\it inequivalent} translation  $\vec T_{m'n'}$ (in the same star) are equal.
This occurs only at discrete angles $\theta_{mn}$ that can be indexed by the two integers
\begin{eqnarray}
e^{i \theta_{mn}} = \frac{m e^{-i \pi/6} + n e^{i \pi / 6}}{n e^{-i \pi/6} + m e^{i \pi / 6}}
\end{eqnarray}
Small angle rotations have large $m$ and $n=m+1$. These small angle faults describe large period superlattices where the atomic
registry can be regarded as evolving smoothly between widely separated regions with locally Bernal-like and $AA$ like stacking.
Complementary structures with large $m$ and $n=1$ correspond to small angular deviations from  the $60^\circ$ rotated
structure. Since the combination of a $60^\circ$ degree rotation and a translation by a nearest neighbor bond vector is a
symmetry operation of the honeycomb lattice, a commensurate rotation near $60^\circ$ can be regarded as the superposition of a
small angle rotation and a nonprimitive translation.  The primitive translations $\vec A$ of a general $(m,n)$ commensuration
supercell are
\begin{eqnarray}
\left( \begin{array}{c}
 \vec A_1\\
  \vec A_2 \\
\end{array} \right) = \left(
\begin{array}{cc}
  m & n \\
 -n  & m+n \\
\end{array}
\right) \left( \begin{array}{c}
 \vec a_1\\
  \vec a_2 \\
\end{array} \right)
\end{eqnarray}
with length $|\vec A| = \sqrt{m^2 + n^2 + mn}$.

Commensuration pairs at angles $\theta$ and $\bar \theta = 60^\circ - \theta$ are related. The simplest example of such a pair
occurs trivially for $\theta = 0$ and $\bar \theta = 60^\circ$ which correspond to the smallest possible unit cells with $AB$
(Bernal) stacking and $AA$ (perfectly eclipsed) stacking. The unit cells of these structures have the same area but they have
different sublattice symmetries. Importantly, all commensurate rotations share this property: they occur in partners where the
sum of the rotation angles is $60^\circ$ and their unit cells have the same area. The commensuration indices $(m,n)$ and $(\bar
m, \bar n)$ of the partners are related
\begin{eqnarray}
\left( \begin{array}{c}
  \bar m \\
   \bar n \\
\end{array} \right) = \left(
\begin{array}{cc}
  -1 & 1 \\
 2  & 1 \\
\end{array}
\right) \left( \begin{array}{c}
  m\\
   n \\
\end{array} \right)
\end{eqnarray}
eliminating common divisors by 3 from the result. Figure 1 illustrates this situation where the structure in the left panel
corresponds to $(m,n) = (1,3)$, $\theta = 32.204^\circ$ and on the right $(\bar m, \bar n) = (2,5)$, $\bar \theta =
27.796^\circ$. Partner commensurations can also be transformed into each other by a translation $\vec \Delta$ at a {\it single}
rotation angle $\theta$ demonstrating the invariance of the primitive cell area. The structure at $\theta = \bar \theta =
30^\circ$ is its own commensuration partner and corresponds to an elementary two dimensional quasicrystalline lattice. Note
also that the form of Eqn. 1 demonstrates that the indices $(m,n)$ generally provide a more useful specification of the
structure than the fault angle $\theta$. Indeed nearby rotation angles can have very different fault indices and therefore
describe crystalline structures with vastly different periods and different physical properties. A plot of the commensuration
periods $|\vec A|$ as a function of rotation angle $\theta$ shows a complex distribution of allowed periods which is bounded
from below. This lower bound has a smooth dependence on $\theta$, diverges as $\theta \rightarrow (0^\circ, 30^\circ,
60^\circ)$ and is symmetric around the self dual state at $30^\circ$.

\begin{figure}
\begin{center}
\includegraphics[angle=0,width=100mm,bb=0 0 500 300]{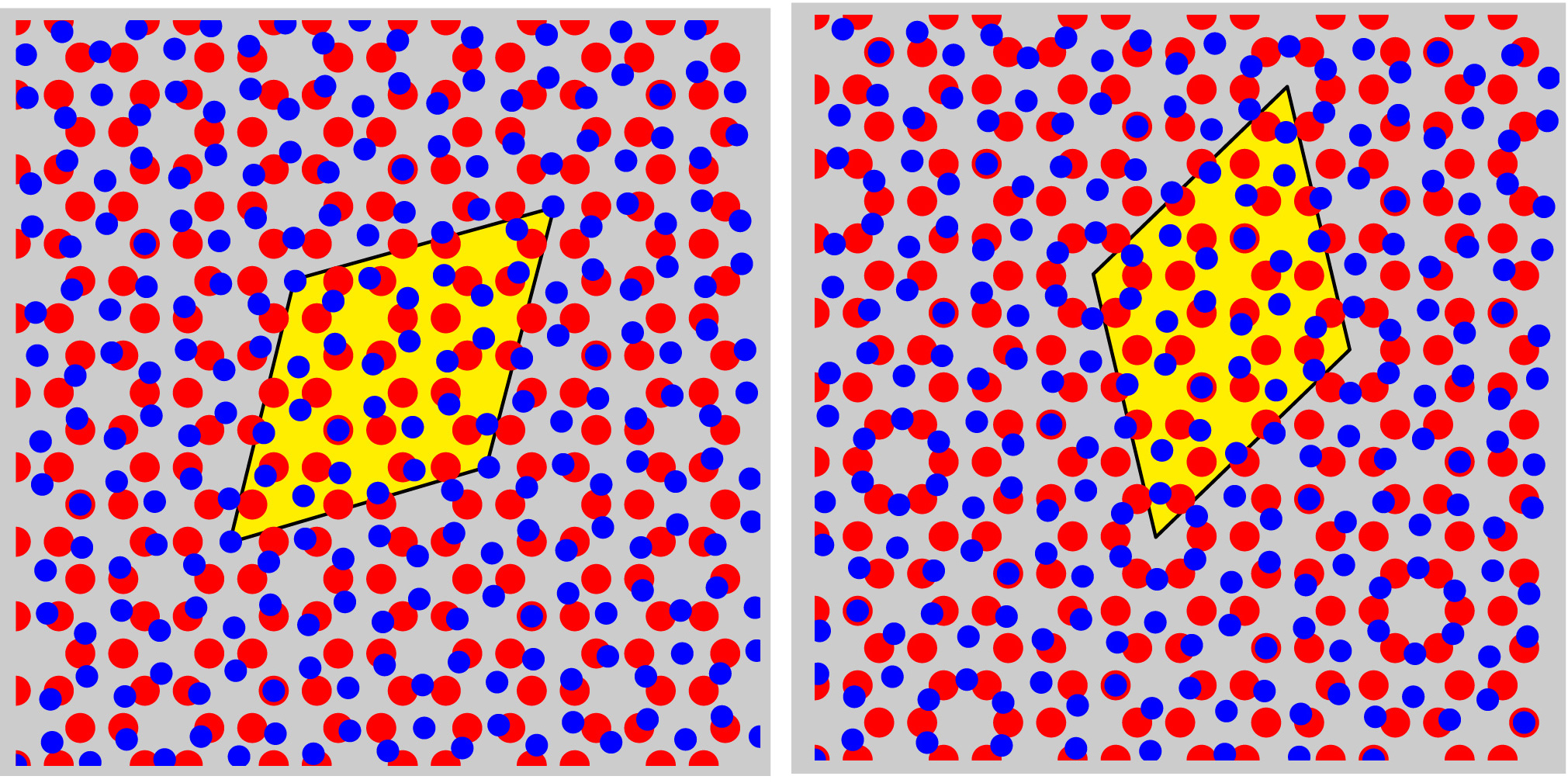}
\caption{\label{lattice}  Two lattice structures for rotationally faulted graphene bilayers at complementary rotation angles.
Red and blue dots denote atomic positions in different layers. The highlighted rhombus is a primitive commensuration cell.  The
figure compares the
  stacking patterns for commensuration pairs that are related by $\bar{\theta} =60^\circ - \theta$. The structures are
  (left)  $\theta=32.204^\circ$ ($m=1,n=3$)  and (right) $27.796^\circ$ ($m=2,n=5$).  The commensuration cells are the same for the partner structures but the point symmetry is
  different.}
\end{center}
\end{figure}

Commensuration partners are distinguished by their {\it sublattice exchange parity}. A commensuration is sublattice exchange
``even" if the commensuration cell contains an $A$ and a $B$  sublattice site in each layer that are coincident with atomic
sites in the neighboring layer. A commensuration is sublattice exchange ``odd" if only one sublattice site in the
commensuration cell is eclipsed. (Fixing the rotation center of the twist at an atom site guarantees that there will be at
least one coincident site.) The sublattice exchange parity can be deduced from the translation indices $(m,n)$. It is
convenient to label the eclipsed sites at the origin as the $A$ sublattice, a nearest neighbor bond vector $\vec \tau$ and its
partner in the rotated layer $\tau'$. Then the condition for a second coincident site on the $B$ sublattice is
\begin{eqnarray}
\vec T + \vec \tau = \vec T' + \vec \tau'
\end{eqnarray}
for some possible choice of $\vec T(\vec T')$ in the set of lattice translations in the reference(rotated) layers. Since the
$\vec T$'s are both lattice translations, this requires integer $(p,q)$ solutions to
\begin{eqnarray}
e^{i \theta_{mn}}
 = \frac{ 1 + \sqrt{3} \left(p e^{i \pi/6} + q e^{-i \pi/6} \right)}{ 1 + \sqrt{3} \left(p e^{-i \pi/6} + q
e^{i \pi/6} \right)} =  \frac{ m e^{i \pi/6} + n e^{-i \pi/6}}{ m e^{-i \pi/6} + n e^{i \pi/6} }
\end{eqnarray}
which can be expressed
\begin{eqnarray}
p = \frac{m - n + 3mq}{3n}
\end{eqnarray}
and has integer solutions only when $m-n$ is divisible by $3$.  When this is statisfied the coincident sites occur at special
high symmetry points in the cell $\vec A_c = \pm \vec A_{mn}/3$ (with only one sign per structure) and correspond to high
symmetry positions along the diagonal of the rhombus shown in the right hand panel of Figure 1.  When $m-n$ is not divisible by
$3$ the only coincident site occurs at the center of rotation and its supercell translates.

\subsection{Reciprocal Space}

Similar considerations apply to the momentum space representation of the twisted bilayer for which Figure 2 gives a map
illustrating the structure of its reciprocal space. The reciprocal lattice of the commensuration supercell can be treated as a
conventional triangular lattice spanned by two primitive vectors $2 \pi (\hat e_z \times \vec A_i)/{\cal A}$ where ${\cal A} =
|\vec A_1 \times \vec A_2|$ is the area of the commensuration supercell and $\hat e_z$ is the layer normal. However, it is
often useful to observe that since the real space lattice translations of the supercell are coincident lattice translations of
each of the layers, its reciprocal space can also be indexed by a reciprocal lattice spanned by momenta with {\it four} integer
indices describing all linear combinations of the primitive reciprocal lattice translations of each of the layers, in the
manner
\begin{eqnarray}
\vec {\cal G}_{p,q,p',q'} = p \vec G_1 + q \vec G_2 + p' \vec G_1' + q' \vec G_2'
\end{eqnarray}
This demonstrates that the primitive $\vec G$'s and $\vec G'$'s and all possible combinations $\vec G + \vec G'$ are in the
reciprocal lattice of the faulted structure. The smallest nonzero combinations of these vectors have length $4 \pi/\sqrt{3 (m^2
+ mn +n^2)}$ and span the first star of reciprocal lattice vectors of the commensuration supercell.

\begin{figure}
\begin{center}
\includegraphics[angle=0,width=40mm,bb=0 0 300 300]{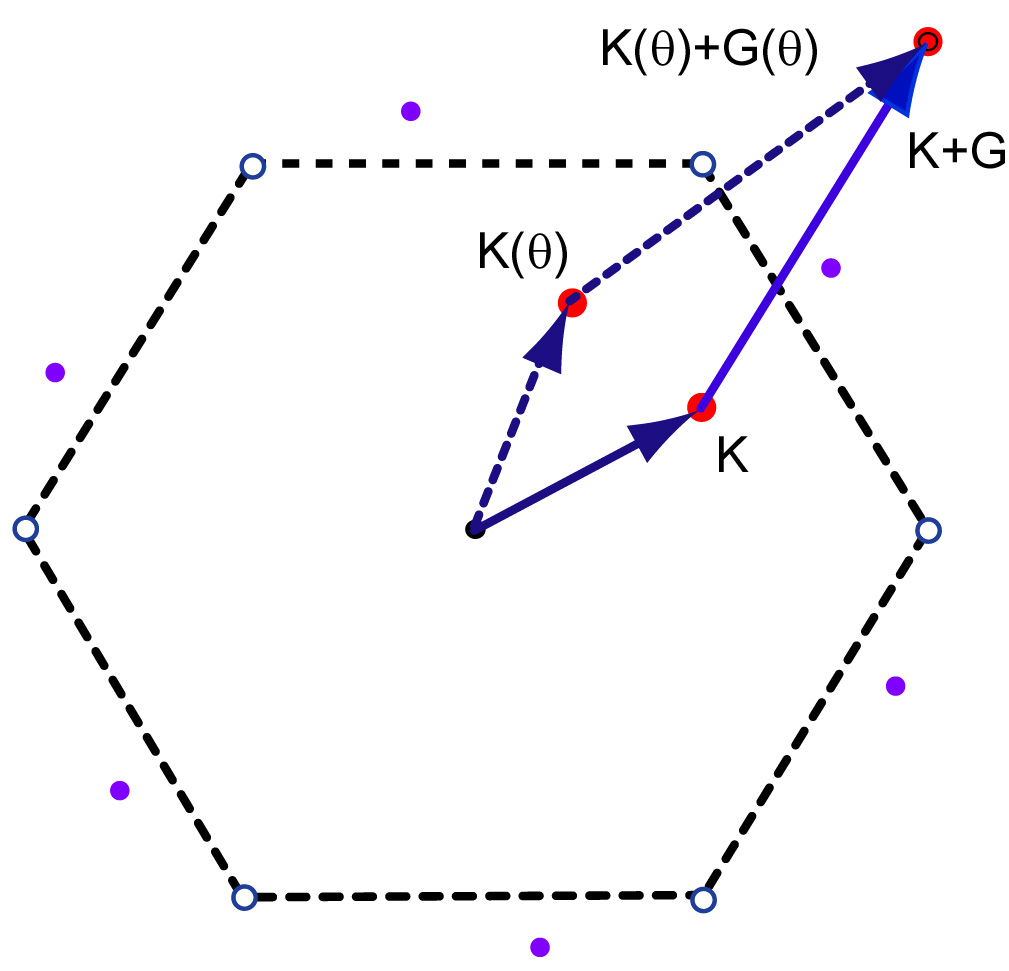}
\end{center}
\caption{\label{recipspace}  A reciprocal space map for a twisted graphene bilayer illustrating the rotation of the first star
of reciprocal lattice vectors (open dots) to a star of rotated reciprocal lattice vectors (filled blue dots), and a
corresponding rotation of the zone corner $K$ points (red dots). The offset points $K$ and $K(\theta)$ become coincident in the
extended zone after translations by a particular pair of reciprocal lattice vectors.}
\end{figure}

A critical question is whether the momentum offset $\vec K(\theta)- \vec K$ or $K(\theta) - \vec K'$ are {\it also} in the
reciprocal lattice of the commensuration cell. For the former situation this is the question of whether
\begin{eqnarray}
\vec K(\theta) - \vec K = \vec {\cal G}_{p,q,p',q'} = p \vec G_1 + q \vec G_2 + p' \vec G_1' + q' \vec G_2' \, (?)
\end{eqnarray}
for some choice of integers $(p,q,p',q')$. Representing these two dimensional vectors by complex numbers one finds that Eqn. 9
can be expressed
\begin{eqnarray}
e^{i \theta_{mn}} = \frac{ 1 + \sqrt{3} \left(p e^{i \pi/6} + q e^{-i \pi/6} \right)}{ 1 + \sqrt{3} \left(p' e^{-i \pi/6} + q'
e^{i \pi/6} \right)}
\end{eqnarray}
where $\theta_{mn}$ is given by Eqn. 2. Nontrivial solutions invert the indices $p'=q$ and $q'=p$ and lead to the matching
condition
\begin{eqnarray}
p = \frac{m-n - 3mq}{3n}
\end{eqnarray}
Thus $\vec K(\theta) - \vec K$ is in the reciprocal lattice only for {\it supercommensurate} structures where $m-n$ is a
multiple of 3. Eqn. 11 is identical to Eqn. 7 that identifies the even sublattice exchange commensurations, so that sublattice
``even" structures always allow intravalley interlayer coupling. For example, when $(m,n) = (2,5)$  we have $\theta =
27.796^{\circ}$ and the first integer solutions to Eqn. 10 occur for $q=3$ for which $p=1$ and $(p',q') = (3,1)$. The existence
of this solution implies that these $K$ points are coincident in the extended zone after translations by $p \vec G_1 + q \vec
G_2$ and $p' \vec G'_1 + q \vec G'_2$ as illustrated in the right panel of Fig. 3.

One can also ask about the possibility of commensurability for intervalley momentum transfer $K(\theta) - K'$, namely
\begin{eqnarray}
\vec K(\theta) - \vec K' = \vec {\cal G}_{p,q,p',q'} = p \vec G_1 + q \vec G_2 + p' \vec G_1' + q' \vec G_2' \, (?)
\end{eqnarray}
Following a similar line of analysis one finds a different set of commensurability conditions
\begin{eqnarray}
p &=& \frac{m(q+1) -nq}{m+2n} \nonumber\\
p' &=&  \frac{nq -m(1+q)}{m+2n} \nonumber\\
q' &=& \frac{(2m+n)q + m}{m+2n}
\end{eqnarray}
Thus for example, $m=1$, $n=3$ gives a rotation angle $\theta = 32.204^\circ$ which has its first integer solution when $\tilde
q = q = 4$, giving $(p,q) = (-1,4)$ and $(p',q') = (1,3)$. Notice the asymmetry between the values of $(p,q)$ and $(p',q')$:
the scattering between inequivalent Dirac cones requires {\it different} umklapp terms when indexed to the individual
reciprocal lattices of the two layers.  The indices would be reversed by considering $K \rightarrow K'(\theta)$ couplings. This
matching rule implies that $\vec K(\theta)$ and $\vec K'$ are coincident in after translations by $p \vec G_1 + q \vec G_2$ and
$p' \vec G'_1 + q \vec G'_2$ as illustrated in the left panel of Fig. 3.

\begin{figure}
\includegraphics[angle=0,width=150mm,bb=0 0 600 250]{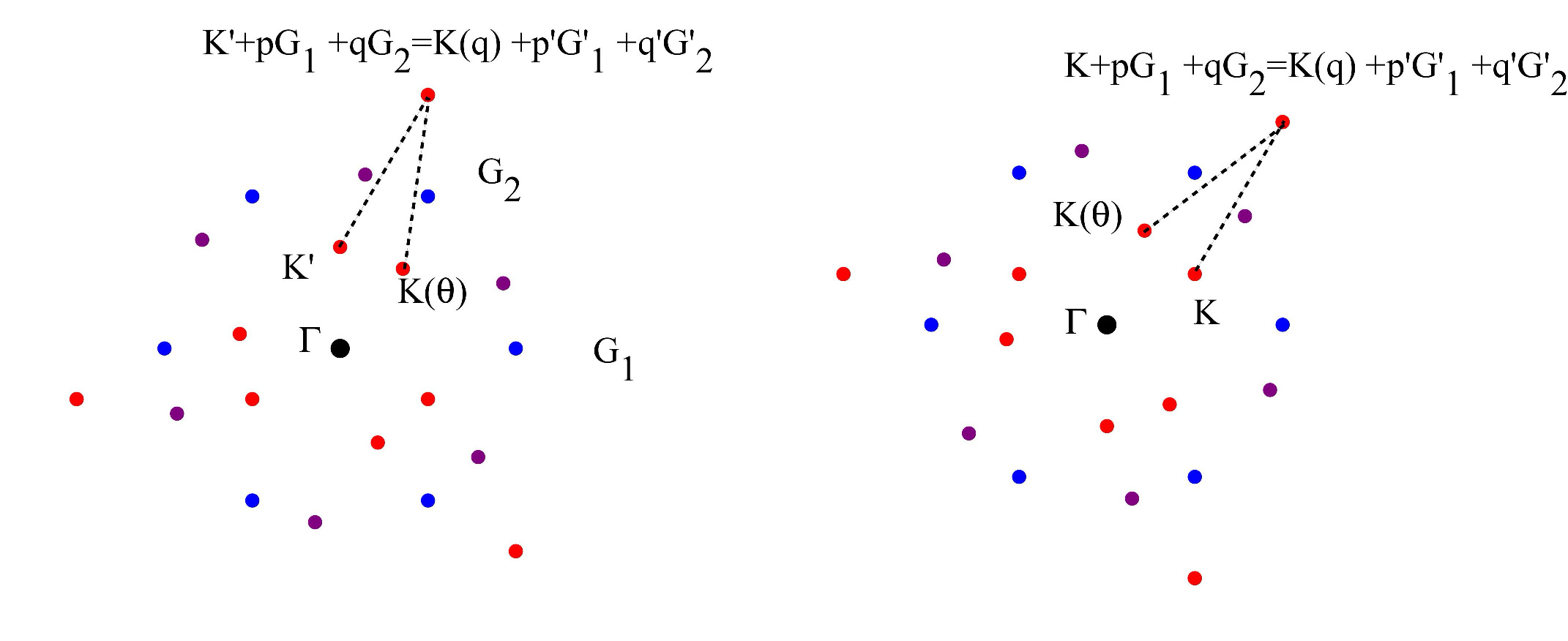}
\caption{\label{recipspace2}  Commensurability conditions in reciprocal space for partner commensurations at (left) $\theta =
21.787^\circ$ ($(m,n) = 1,3$) and (right) $\theta = 38.213^\circ$ ($(m,n) = (2,5)$). In the left panel the lattice structure is
odd under sublattice exchange and the offset $K(\theta) - K'$ (red dots) is in the reciprocal lattice of the commensuration
cell. Translation by particular layer reciprocal lattice vectors (spanned by the layer reciprocal lattice vectors and their
rotated counterparts, shown as the blue and violet points, respectively)  brings these two momenta into coincidence in the
extended zone. In the right panel the lattice structure is even under sublattice exchange and the offset $K(\theta) - K$ is in
the reciprocal lattice of the commensuration cell. Translation by different layer reciprocal lattice vectors brings these two
momenta into coincidence in the extended zone. These two commensuration conditions are complementary and mutually exclusive.}
\end{figure}

One can prove that Eqns. 9 and 12 cannot be simultaneously satisfied for a common rotation angle. For example if $m=3 \mu + 1$
and $n = 3 \nu + 1$ then $m-n$ is a multiple of 3 and intravalley couplings are in the reciprocal lattice. In this situation
Eqn. 13 requires that
\begin{eqnarray}
p = \frac{3(\mu q - \nu q + \mu) + 1}{3(\mu + 2 \nu +1)}
\end{eqnarray}
which is impossible since the numerator is never divisible by 3 so that intervalley coupling is excluded. On the other hand,
when $m=3 \mu \pm 1$ and $\nu = 3 \nu \mp 1$ intravalley are excluded and
\begin{eqnarray}
p = \frac{(3(\mu - \nu)q + \mu) + 2q +1}{3(\mu + 2 \nu) \mp 1 }
\end{eqnarray}
which can be satisfied for integer $p$ by appropriate choice of integer $q$ so that intervalley coupling is allowed. These two
possibilities are thus complementary and mutually exclusive: one or the other must occur if the rotation is commensurate. Using
the indexing rule Eqn. 4, one can easily show that partner commensurations realize complementary commensuration conditions: one
member admits only the intravalley interlayer coupling while the other allows only the analogous intervalley scattering.  Thus
the valley structure of the interlayer couplings are specified by the sublattice exchange symmetry of the structure.

\section{Layer Decoupling by Rotational Mismatch}

Early work on the electronic properties of twisted graphene bilayers recognized that small angle rotational faults are
inevitably described by large period commensuration cells that make atomistic calculations impractical. Instead it is useful to
develop a long wavelength description that captures the effect of rotation on the low energy electronic structure. The
essential physics in this treatment is the momentum offset of the Dirac nodes produced by the rotation  \cite{JMLdS}  as
illustrated in Fig. 2.

The starting point of the continuum theory is the long wavelength theory appropriate to {\it single layer} graphene. The
effective mass theory for electrons in each valley introduces two Dirac Hamiltonians for their smoothly varying pseudo-spinor
fields
\begin{eqnarray}
H_K = - i \hbar v_F  \sigma \cdot \nabla; \,H_{K'} = \sigma_y H_K \sigma_y
\end{eqnarray}
where the $\sigma$'s are $2\times 2$ Pauli matrices acting on the sublattice amplitudes. A small angle relative rotation of the
crystallographic axes in the of the two layers offsets the crystal momenta of their closest Dirac nodes by $\Delta K = 2K \sin
(\theta/2)$. This can be described by a pair of layer-polarized Dirac Hamiltonians parameterized by the momentum offset $\Delta
\vec K = \vec K(\theta) - \vec K$ in the manner
\begin{eqnarray}
H_K &=&   \hbar v_F  \sigma \cdot (-i \nabla  - \frac{\Delta \vec K}{2})\nonumber\\
H_{K(\theta)} &=&  \hbar v_F  \sigma^\theta \cdot (-i \nabla  +  \frac{\Delta \vec K}{2})
\end{eqnarray}
where $\sigma_\mu^\theta = \exp(i \sigma_z \theta/2) \sigma_\mu \exp(-i \sigma_z \theta/2)$ because of the relative rotation of
the two layers.

Electrons in neighboring layers are coupled by a $\theta$-dependent interlayer coupling amplitude projected into the pseudospin
basis. For small angle rotations these interlayer amplitudes vary smoothly in real space and one can focus on their lowest
Fourier components.  In the theory of Lopes dos Santos {\it et al.} \cite{JMLdS} the offset momentum $\Delta \vec K$ is not in
the reciprocal lattice of the commensuration cell and these authors focus on the three momenta $\vec {\cal G}_i$ that leave the
offset $|\vec K(\theta)  - \vec K - \vec {\cal G}_i|$ invariant. These momenta can be expressed in terms of the offset $\Delta
K$: in complex notation they  are ${\cal G}_i = (0,{\cal G}_1 = \sqrt{3}e^{i \pi/6} \Delta K,{\cal G}_1 + {\cal G}_2 = \sqrt{3}
e^{-i \pi/6} \Delta K)$.  In the pseudospin basis, the interlayer coupling for each of these momentum transfers is
characterized by a $2 \times 2$ matrix-valued tunneling coefficient $T({\cal G})$ whose elements have been estimated
numerically using a tight binding model. This yields in the small angle limit
\begin{eqnarray}
T({\cal G}=0)  =  \tilde t_{\perp} \left(%
\begin{array}{cc}
  1 & 1 \\
  1 & 1 \\
\end{array}%
\right) ; \,\,\,    T({\cal G}= -{\cal G}_1) = \tilde t_{\perp} \left(%
\begin{array}{cc}
  z & 1 \\
  \bar z & z \\
\end{array}%
\right)  \nonumber\\
\,\,\,\,\,\,  T({\cal G} = -({\cal G}_1 + {\cal G}_2) )= \tilde t_{\perp} \left(%
\begin{array}{cc}
  \bar z & 1 \\
  z & \bar z \\
\end{array}%
\right)
\end{eqnarray}
where $z=e^{2 \pi i /3}$, $\bar z = e^{-2 \pi i/3}$ and $\tilde t_{\perp} \sim 0.11 \, {\rm eV}$,  approximately independent of
the supercell period.

The asymmetry in the set of selected $\vec {\cal G}$'s appearing in Eqn. 18 occurs because of the choice of the reference
valley for the long wavelength expansion. Nevertheless, this approach explicitly preserves the threefold rotational symmetry of
the commensuration cell. This is seen most clearly by observing that the $T$ matrices are off diagonal operators in the layer
degree of freedom and one may therefore arbitrarily ``shift" the interlayer coupling  momenta by a layer-dependent $U(1)$ gauge
transformation. In particular the gauge shift $e^{-i \Delta \vec K \cdot \vec r}$ in the rotated layer brings the two Dirac
nodes into coincidence and shifts the three momentum transfers so that they form the three arms of a star  generated by
$Q=-\Delta K$ and its $\pm 2 \pi/3$-rotated partners. The negates of these  three momenta occur in the reciprocal amplitudes
describing the reverse tunneling processes. Thus the expansion about a {\it single} zone corner point preserves the full three
fold symmetry of the commensuration cell, as required

The essential features of this theory are (1) the existence of a crystal momentum offset due to the rotational fault, (2) the
coupling of plane wave states in one layer to a triad of plane wave states in its neighbor and (3) the existence of a ${\cal
G}=0$ term in the effective interlayer tunneling Hamiltonian. Feature (1) suggests that at sufficiently low energy the effect
of the interlayer coupling can be treated perturbatively in the dimensionless coupling parameter $\Gamma = \tilde
t_{\perp}/\hbar v_F \Delta K$. Feature (2) implies the perturbative effects of this coupling {\it vanish by symmetry} precisely
at $E=0$  so that the coupled system preserves the Dirac nodes of its two  (decoupled) layers. Perturbative effects of the
coupling arise at linear order in the momentum differences $\vec k \pm \Delta \vec K/2$ and can be interpreted as a twist
dependent renormalization of the Fermi velocity
\begin{eqnarray}
\frac{v^*_F}{v_F} = 1 - 9 \left( \frac{\tilde t_{\perp}}{\hbar v_F \Delta K} \right)^2
\end{eqnarray}
Equation 19 needs to be applied with care since it breaks down both in the limit of small rotation angles due to a failure of
the perturbation theory when $\Delta K \rightarrow 0$  and at large rotation angles when commensuration effects, neglected in
this treatment, can intervene. Finally, feature (3) indicates that electron states in the two layers that have the {\it same}
crystal momentum modulo $\vec {\cal G}$ are  coupled through the interlayer Hamiltonian.  In the low energy theory the
layer-polarized Dirac cones degenerate in the planes that bisect lines connecting their nodes (above the crossover energy
$\hbar v_F \Delta K/2$) and one expects the strongest interlayer mixing to occur in these planes. There are three such planes
that bisect the lines along $\Delta K$ and its $\pm 2 \pi/3$ rotated counterparts. The onset of this mixing is associated with
a change of topology of the bilayer bands, connecting a low energy sector with layer-decoupled Dirac cones to higher energy
layer-coherent hyperbolic bands. In the lowest band this transition is associated with a saddle point in the electronic
spectrum and a logarithmic van Hove singularity in the two dimensional density of states \cite{vanHove}.

\section{Atomistic Calculations}

The novel physics of rotationally-induced layer decoupling has stimulated theoretical work by several groups to explore this
effect using various atomistic models. Ab initio calculations have been carried out for misaligned bilayer supercells
containing up to $\sim 500$ atoms ($\theta \sim 5^\circ$) while tight binding methods have allowed workers to access larger
systems of up to $15000$ atoms \cite{haas,Latil ,GTdL ,Shallcross ,turbo}. These studies have examined the Fermi velocity
renormalization, the form of the low energy electronic spectrum near the $K$ points and the spatial modulation of the
electronic charge density.

Much of the ab-initio work has understandably focused on the shortest period twisted structures, e.g. $\sqrt{7} \times
\sqrt{7}$ and $\sqrt{13} \times \sqrt{13}$ commensurations \cite{haas,Shallcross }. Calculations on these systems generally
confirm a suppression of the interlayer coupling scale and a Fermi velocity near the $K$ point which is essentially
indistinguishable from that of single layer graphene. The most thorough investigation of the Fermi velocity renormalization has
been given by de Laissardi$\grave{\rm e}$re {\it et al.} \cite{GTdL } who suggest that the rotational faults are characterized
by three different velocity renormalization regimes, determined by the fault angle: (a) $ 15^\circ < \theta < 30^\circ$ where
the Fermi velocity is essentially the same as for single layer graphene, (b) $3^\circ < \theta < 15^\circ$ where a downward
renormalization is found, well described by the perturbation theory of \cite{JMLdS}, and a low angle regime $\theta < 3^\circ$
where the low energy bands are flattened and not described by the perturbative treatment. The small renormalization in the
large fault angle regime (a) is at least qualitatively consistent with the continuum theory since the renormalization occurs
via a virtual mixing of low energy states with states separated by an energy barrier $\hbar v_F \Delta K$.  The breakdown of
the perturbation theory for sufficiently small angle faults is similarly understandable since it involves an expansion in
$\tilde t_{\perp}/\hbar v_F \Delta K$. Surprisingly, in this low angle regime de Laissardi$\grave{\rm e}$re {\it et al.} also
report a pronounced spatial modulation of the low energy eigenstates that tends to localize their charge densities in spatial
regions locally characterized by ``AA" stacking, suggesting some form of multiband physics that is not captured by the
truncated continuum model.  The accuracy of the perturbation theory in the intermediate regime has been further questioned by
the density functional calculations of Shallcross {\it et al.} \cite{Shallcross } who find that the bilayer $v_F$ is nearly
equal to that of single layer graphene down to smallest angles ($\sim 9^\circ$) they were able to study.

The calculations by Shallcross {\it et al.} \cite {Shallcross } also reveal features in the electronic spectra near the $K$
points that are not captured by the primitive continuum theory. Significantly, close to the zone corners the electronic bands
are {\it not} linear but instead they are mixed, which requires an interlayer mass operator in the low energy Hamiltonian.
Interestingly the spectral structure, and therefore the matrix structure of this mass term, is {\it different} for partner
commensurations and therefore it cannot be determined solely by the size of the commensuration cell. The scale of the mixing is
nevertheless small relative to its value for Bernal bilayers, e.g. the mixing scale for the $\theta = 30^\circ \pm 8.213^\circ$
structure is $\approx 7 \, {\rm meV}$ compared to $\approx 0.2 \, {\rm eV}$ for Bernal stacking
\cite{McCFal,koshino,biasedbilayer}. Further, over the range of structures they studied this mass scale appears to a rapidly
decreasing function of the commensuration cell period. But the existence of this mass matrix in the low energy theory presents
a significant challenge to the interpretation of the electronic states even at energies {\it above} the mass scale. Notably, in
order to match smoothly to these low energy eigenstates the bilayer eigenstates at higher energy must be (nearly) {\it equal
weight} states coherently mixed between the two layers instead of the layer-polarized eigenstates that one would infer from the
momentum space structure of the continuum theory .

\section{Second Generation Continuum Theories}

There has been progress in the development of new long wavelength models that extend the physics identified in the original
continuum formulation \cite{JMLdS}. These theories examine the effects of lattice commensuration \cite{ejmrc} and of multi-band
mixing \cite{bismac1} on the low energy electronic structure. The former turn out to be most important for large angle faults
while the latter are critical to the physics at small rotation angles. The new models are also formulated as continuum theories
in order to circumvent the technical difficulty posed by fully microscopic atomistic treatments of large commensuration cells.
Concurrently there has been an effort to distill the original continuum model to a simpler effective two band model
\cite{saddledeGail,saddleLL} in an effort to explore the effects of the novel band topology on the orbital quantization of its
electronic states in a perpendicular magnetic field. We refer to all these new models as ``second generation" continuum
theories.

\subsection{Interlayer Matrix Elements}

A microscopic theory of the interlayer coupling can be formulated in the basis of Bloch orbitals
\begin{eqnarray}
\psi_\alpha (\vec k) = \frac{1}{\sqrt{N}} \sum_{\vec T} e^{i \vec k \cdot (\vec T + \vec \tau_\alpha)} \phi_\alpha (\vec T)
\end{eqnarray}
where $\phi_{\alpha = (A,B)}$ are orbitals centered at positions $\vec T + \vec \tau_{\alpha}$ and $\vec T$ is a lattice
translation. In this basis the interlayer Hamiltonian is
\begin{eqnarray}
\langle \psi_\beta(\vec k') | {\cal H} | \psi_\alpha(\vec k) \rangle = \frac{1}{N} \sum_{\vec T, \vec T'} \, e^{-i \vec k \cdot
(\vec T' + \vec \tau_\beta)} \langle \phi_\beta (\vec T') | {\cal H} | \phi_\alpha (\vec T) \rangle e^{i \vec k \cdot (\vec T +
\vec \tau_\alpha)}
\end{eqnarray}
Assuming that the inter-site tunneling amplitude depends on the layer-projected difference coordinate, the matrix element can
be expressed
\begin{eqnarray}
\langle \phi_\beta (\vec T') | {\cal H} | \phi_\alpha (\vec T) \rangle = \frac{1}{(2 \pi)^2} \int \, d^2 q \, f(\vec q) \, e^{i
\vec q \cdot (\vec T' + \vec \tau'_\beta - \vec T - \vec \tau_\alpha)}
\end{eqnarray}
Carrying out the lattice sums in Eqn. 21 and expressing the momenta in terms of their differences from the respective zone
corners, $\vec k = \vec K + \vec q$,  one obtains an expression for the interlayer tunneling amplitude in terms of sums over
the reciprocal lattices of the reference ($\vec G$) and rotated ($\vec G'$) layers
\begin{eqnarray}
{\cal T}_{\beta \alpha} (\vec q',\vec q) = \frac{1}{\cal A} \sum_{\vec G, \vec G'} \, f(\vec q + \vec K + \vec G) \, e^{i \vec
G' \cdot
\vec \tau_\beta} e^{-i \vec  G \cdot \tau_\alpha} \, \nonumber\\
 \,\,\,\,\,\,\,\,\,\,\,\,\,\, \delta(\vec q' - \vec q +  \Delta K  + \vec G' -
\vec G)
\end{eqnarray}
where ${\cal A}$ is the area of the unit cell.

When  $q \ll G$ Eqn. 23 describes two distinct types of interlayer tunneling processes:  (1) Direct interlayer terms conserve
the {\it crystal momentum} $\vec k$ and occur when  $\Delta K = |\vec K(\theta) - \vec K + \vec G' - \vec G|$. (Note that this
occurs for $\vec G = \vec G' = 0$ and for all boosts by the reciprocal lattice vectors $\vec {\cal G} = \vec G' - \vec G$ that
symmetrically shift the initial and final states to nearby valleys.) (2) Indirect interlayer terms conserve the {\it Dirac
momentum} $\vec q$ and occur when $\vec K + \vec G = \vec K(\theta) + \vec G$. The matrix element for this latter process is
dominated by the Fourier amplitude of the tunnelling potential at the first momentum $\vec K + \vec G_c$ in the extended zone
where the zone corner points coincide. Processes (1) and (2) have very different character. In the Dirac language, process (1)
requires $\vec q \neq \vec q'$ and provides a microscopic basis for the continuum formulation of Lopes dos Santos {\it et al.}
\cite{JMLdS}. By contrast process (2) allows (indeed requires) $\vec q = \vec q'$ coupling and in particular it provides a
mechanism for {\it coupling between  the tips of the Dirac cones in neighboring layers}. It can be understood as an interlayer
umklapp process whereby the scattering by a reciprocal lattice vector of the commensuration cell provides precisely the right
momentum to bridge the momentum offset $\Delta K$. The ratio of the amplitudes for the indirect and direct couplings is
approximately $f(|\vec K + \vec G|)/f(|\vec K|)$ so that the indirect term is generally weaker than the direct term.

\subsection{Superlattice Commensuration Effects}

As discussed in Section 2.2 either $K \rightarrow  K(\theta)$ or $K' \rightarrow K(\theta)$ couplings are in the reciprocal
lattice of the commensuration cell for a faulted bilayer, depending on the sublattice symmetry, and using Eqn. 23 they are
allowed interlayer tunnelling processes. The low energy theory is fundamentally changed by these terms since they introduce an
interlayer mass operator in the long wavelength Hamiltonian. Interestingly, the analytic structure of this mass matrix is
determined solely by the sublattice symmetry of the commensuration. Thus one can define two complementary families of
commensurate faults where all members of a common family have a common form for their low energy Hamiltonians. The energy scale
of this mass operator depends on the period of the commensuration, and  it is largest for low order commensurate rotations. The
primitive stacked structures with $AB$ and $AA$ stacking are parent structures for this family behavior which give prototypical
examples for the interlayer mixing possible for generic commensurate bilayers.

The commensuration physics for these systems can be understood most easily by explicitly writing the layer Bloch states in real
space in the ``first star" approximation that retains only the three reciprocal lattice vectors that keep the combination $\vec
K + \vec G$ to the first star of $K$ points
\begin{eqnarray}
\Psi(\vec r) = \sum_\alpha \, \Phi_\alpha (\vec r) u_\alpha (\vec r);  \Phi_\alpha(\vec r) = \frac{1}{\sqrt{3}} \sum_{m=1}^3 \,
e^{i \vec K_m \cdot (\vec r - \vec \tau_\alpha)}
\end{eqnarray}
The coupling between layers is a functional of the Bloch fields $\Psi(\vec r)$ and its properties are captured by the local
functional
\begin{eqnarray}
U=\frac{1}{2} \, \int \, d^2r \, T_{\ell} (\vec r) |\Psi_1(\vec r) - \Psi_2(\vec r)|^2
\end{eqnarray}
where $T_{\ell}$ is a real modulated supercell-periodic function arising from the lattice structure of the commensuration cell.
The coupling function acts to correlate the amplitudes and phases of the $\Psi$'s in the neighboring layers.  The purely {\it
local} coupling between layers in Eqn. 25 can be readily generalized to describe interlayer coupling with a finite range
without substantially changing the physics. Equation 25 describes a coupled mode theory where the full Bloch waves $\Psi$ of
the two layers (importantly these are {\it not} the Dirac envelope functions $u_\alpha(\vec r)$) are coupled by through a local
spatially modulated potential.

Although the exact form of the coupling function ${\cal T}_\ell$ is unknown its important properties are constrained by
symmetry: it is a real supercell-periodic function with local maximum near aligned sites of the two layers and with minima for
regions where atoms in neighboring layers are out of registry. A useful analytic model satisfying these constraints can be
constructed from the elementary density waves in each of the layers
\begin{eqnarray}
n_{\mu=1,2} (\vec r) = \sum_{m \in [1]} \, \sum_{\alpha = A,B} e^{i \vec G_{\mu,m} \cdot (\vec r - \vec \tau_{\mu,\alpha})}
\end{eqnarray}
summed over the first star of reciprocal lattice vectors in the $\mu$-th layer.  Then, a nonlinear functional of the density
fields that satisfies all the symmetry constraints is
\begin{eqnarray}
T_\ell (\vec r) = C_0 e^{C_1(n_1 (\vec r) + n_2(\vec r))}
\end{eqnarray}
The sum of the layer density waves is a real function with the translational symmetry of the commensuration cell and no
shorter. Eqn. 27 is maximized at special positions where the two density functions in each layer are separately maximized
corresponding to aligned atomic sites, and it exhibits exponential suppression in regions where the density waves are out of
registry. This ansatz for the coupling function has some important features. (1) It is a nonlinear function of the primitive
reciprocal lattices vectors of each of the layers so that {\it all} the reciprocal lattice vectors of the commensuration cell
are are represented in an expansion of the exponential in powers of its argument. (2) It is a separable function, constructed
from a product of functions each of which is spanned by the {\it separate} reciprocal lattices of the two layers. (3) It is
parameterized by two constants $C_0$ and $C_1$ which respectively describe the strength and range of the microscopic interlayer
tunneling amplitudes. (Thus for example, very long range hopping is described by a small value of $C_1$.) The two constants
$C_0$ and $C_1$ can be estimated from microscopic theory.

Figure 4 gives a density plot of the local coupling function $T_\ell(\vec r)$ calculated for partner commensurations at $\theta
= 21.787^\circ$ and $\theta = 38.213^\circ$. The former corresponds to a sublattice exchange ``odd" structure and has a
threefold rotational symmetry. Its partner is a sublattice exchange ``even" structure and retains the full sixfold symmetry of
the graphene layer, though on an inflated commensuration supercell.  This illustrates a general property of all ``odd" and
"even" commensurate faults. The patterns shown in this density plot provide a real space image of the interlayer resonance
pattern for a twisted bilayer. Interestingly, for large angle faults, one finds that the appearance of {\it fivefold} resonance
rings (due to rotated misaligned hexagons) is a robust motif in the coupling function. For small angle faults the coupling
function is described instead by the familiar Moire pattern that evolves smoothly between zones locally defined by by $AB$,
$BA$ and $AA$ stacking.

\begin{figure}
\begin{center}
\includegraphics[angle=0,width=80mm,bb=0 0 1200 800]{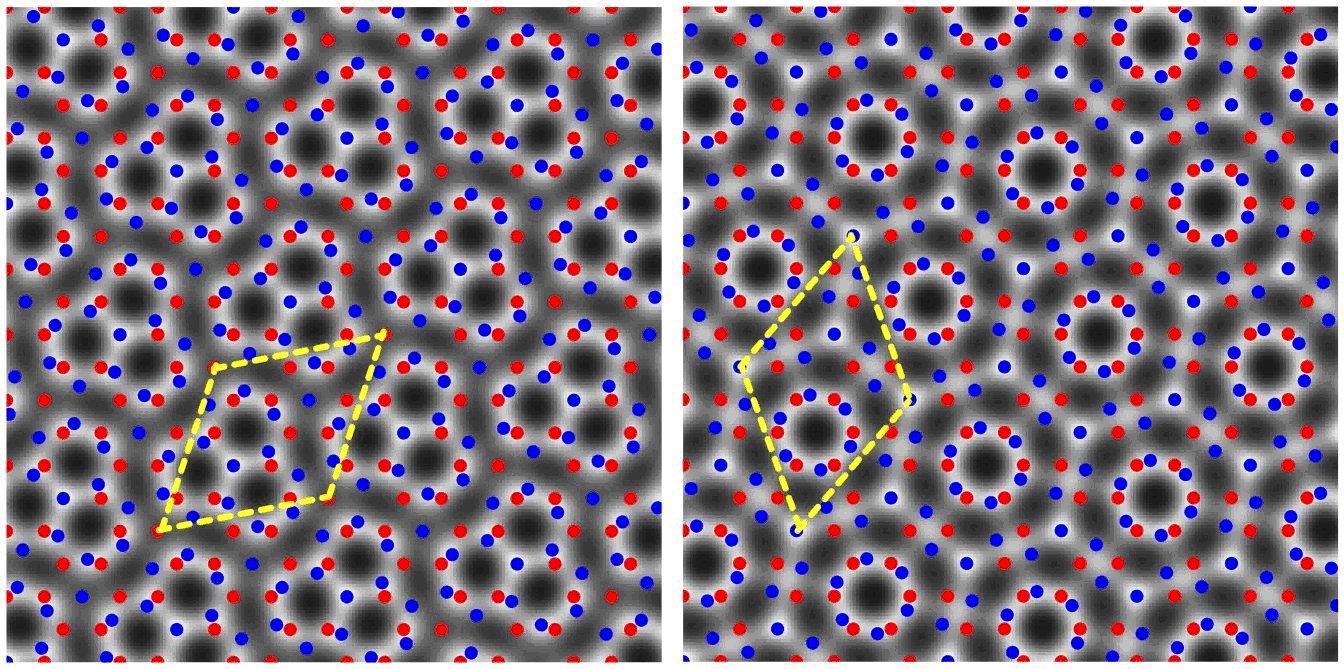}
\end{center}
\caption{\label{hopping}  The spatial dependence of the interlayer coupling function $T_\ell(\vec r)$ of Eqn. 27 is illustrated
by these greyscale plots for commensurate faults at $21.787^\circ$ (left) and $38.213^\circ$ (right). The left structure has
odd sublattice exchange parity,the right structure is even. The left pattern has a threefold rotational symmetry, the right
pattern has the sixfold symmetry of an isolated graphene sheets. Both patterns contain combinations of fivefold resonance rings
that are combined in clusters to form a periodic two dimensional pattern. Adapted from reference \cite{ejmrc}.}
\end{figure}

The couplings between the Dirac fields $u_\alpha$ in neighboring layers are obtained from the cross terms in Eqn. 25 after
integrating out the lattice scale oscillations and are given by the Fourier transform of $T_\ell$ on the reciprocal lattice of
the commensuration cell $t(\vec {\cal G})$. The $\vec {\cal G}= 0$ term describes the crystal momentum-conserving interlayer
couplings discussed in the theory of Lopes dos Santos {\it et al.} \cite{JMLdS}. In addition there are umklapp terms involving
$\vec {\cal G} \neq 0$ terms that express the symmetry allowed couplings between Dirac nodes $K_m \rightarrow K_{m'} (\theta)$.
The geometrical considerations of Section 2.2 require that for any given commensuration there are couplings within two distinct
pairs of Dirac nodes at the corners of their respective Brillouin zones. In the Bloch basis  these matrix elements are spanned
by a $3 \times 3$ matrix of finite momentum scattering amplitudes $\hat V_{\rm ps}$ which, using the threefold rotational
symmetry, takes the form
\begin{eqnarray}
\hat V_{\rm ps} =  \left(%
\begin{array}{ccc}
  V_0 & V_1 & V_2 \\
  V_2 & V_0 & V_1 \\
  V_1 & V_2 & V_0 \\
\end{array}%
\right)
\end{eqnarray}
Here $V_0$ describes the scattering amplitude for a momentum transfer ${\cal G}=|\Delta \vec K|$ coupling the two layers while
$V_1$ and $V_2$ describe scattering amplitudes with larger momentum transfers $\sim \vec G$.   By projecting Eqn. 28 onto the
sublattice (pseudospin) basis, one obtains the $2 \times 2$ interlayer mass matrices $\hat {\cal H}_{\rm int}$ that couple the
Dirac fermions of the two layers. The low energy Hamiltonian for an even bilayer is thus expressed as a $4 \times 4$ matrix
(acting on the two sublattice and two layer degrees of freedom)
\begin{eqnarray}
\hat {\cal H}_{\rm even} = \left( \begin{array}{cc}
  -i \hbar \tilde{v}_F \sigma_1 \cdot \nabla
 & \hat {\cal H}^+_{\rm int} \\
  (\hat {\cal H}^+_{\rm int})^\dagger & -i \hbar \tilde{v}_F \sigma_2 \cdot \nabla
 \\
\end{array} \right)
\end{eqnarray}
while for the odd bilayer
\begin{eqnarray}
\hat {\cal H}_{\rm odd} = \left( \begin{array}{cc}
  -i \hbar \tilde{v}_F \sigma_1 \cdot \nabla
 & \hat {\cal H}^-_{\rm int} \\
  (\hat {\cal H}^-_{\rm int})^\dagger & i \hbar \tilde{v}_F \sigma_2^* \cdot \nabla
 \\
\end{array} \right)
\end{eqnarray}
where $\sigma_n$ are Pauli matrices acting in the sublattice pseudospin basis of the $n-th$ layer and $\tilde{v}_F$ is the
renormalized Fermi velocity. Note that for even parity faults the bilayer Hamiltonian couples nodes of the same chirality,
whereas the odd parity faults introduce coupling between nodes of compensating chirality. In either case the spectrum for
coupled system retains a two-valley character due to the two ways of matching nodes in either family of structures.  The
interlayer mass matrices $\hat {\cal H}^\pm_{\rm int}$ are
\begin{eqnarray}
\hat {\cal H}^+_{\rm int} = {\cal V} e^{i \vartheta} \left(\begin{array}{cc}
  e^{i \varphi/2} & 0 \\
  0 & e^{- i \varphi/2} \\
\end{array} \right) ,\,\,\, \hat {\cal H}^-_{\rm int} = {\cal V} e^{i \vartheta} \left(\begin{array}{cc}
  1 & 0 \\
  0 & 0 \\
\end{array} \right)
\end{eqnarray}
Eqns. 29 and 31 show that for sublattice ``even" faults the mass term involves an $xy$ rotation of its pseudospin through angle
$\varphi$. This angle can not be identified with the rotation angle $\theta$ but it results instead from the interference of
the three complex scattering amplitudes $V_i$. By contrast in Eqns. 30 and 31 one finds that interlayer motion across an ``odd"
fault is mediated by the amplitude on its dominant eclipsed sublattice. In both mass matrices the overall phase of the operator
$\vartheta$ can be removed by a gauge transformation.

Eqn. 31 describes a coupling between Dirac waves in the neighboring layers that persists in the long wavelength $q \rightarrow
0$ limit, qualitatively changing the structure of the low energy spectra. Their effects are illustrated in Figure 5. For
sublattice odd parity faults one pair of coupled bands are gapped on the interaction scale ${\cal V}$ leaving an $E=0$ contact
point between a second pair of quadratically dispersing bands. For the even parity structures the $q=0$ spectrum contains a
pair of coherence-split doublets. These two doublets are the symmetric and antisymmetric combinations of the original single
layer Dirac modes; interestingly the layer-coupled states are topologically required to retain their Dirac character for small
$q $, and disperse linearly away from the points of degeneracy. At finite momentum two branches undergo an avoided crossing
which gaps the spectrum at its charge neutrality points. Thus these structures are generically {\it fully gapped} where the
size of the gap is determined by the pseudospin rotation angle $\varphi$ in Eqn. 31 which allows these branches to hybridize.
This gap degenerates to zero for the special case of an $AA$ stacked structure where $\varphi=0$ by symmetry. In both cases the
residual effect of the mixing at high energy is to introduce a coherence splitting between two linearly dispersing
layer-hybridized bands.

\begin{figure}
\begin{center}
\includegraphics[angle=0,width=80mm,bb=0 0 1600 1600]{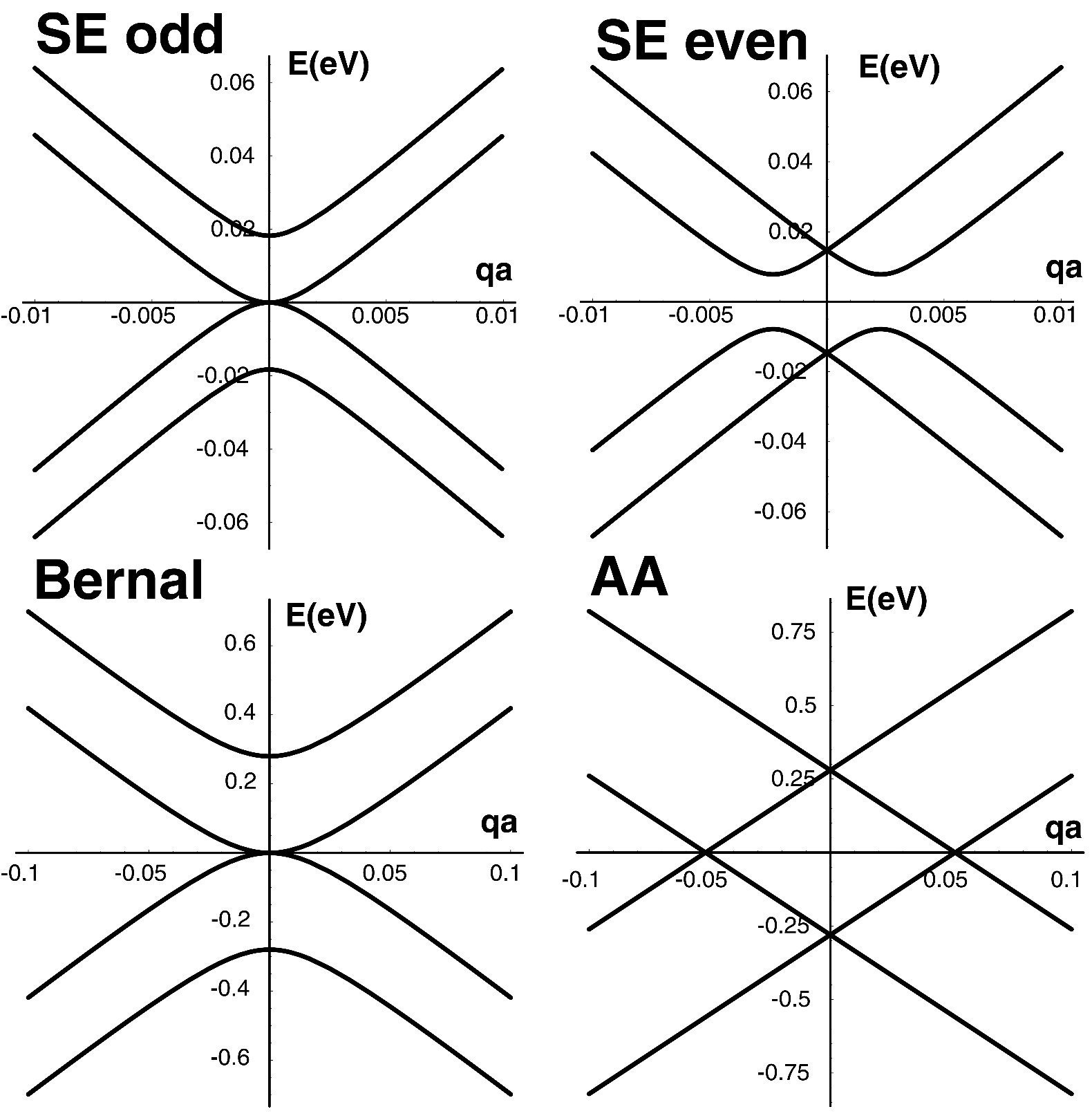}
\end{center}
\caption{\label{massive} The low energe electronic structure for the sublattice exchange odd parity $\theta = 21.787^\circ$
commensuration and the partner even parity $38.213^\circ$ structure are compared with the spectra for the parent Bernal ($AB$)
and $AA$ stacked structures. All odd parity structures, as shown on the left, contain a pair of coherence split massive bands
and a contact point between two quadratic bands at $E=0$. The even parity structures, shown on the right, feature a
bonding/antibonding splitting, with a fully developed gap near the charge neutrality point which degenerates to a gapless state
for the $AA$ stacked structure. Adapted from reference \cite{ejmrc}. }
\end{figure}

Both these behaviors have precise analogs for the limiting cases of Bernal and $AA$ stacked bilayers which can be understood as
the primitive parent structures of these two families. As shown in the lower left panel of Fig. 5 the Bernal spectra exhibit
the mass structure expected for all sublattice exchange odd faults, though on an inflated energy scale ($\approx 0.2 \, {\rm
eV}$) reflecting the full alignment of all sites on a single sublattice. Similarly the primitive $AA$ stacking features a
coherence splitting of its bonding and antibonding layer-coupled states, but without the pseudospin rotation $\varphi$ so that
the spectrum remains gapless and the zero energy states occur on a ring in reciprocal space.

\subsection{Nonlocal Potential Scattering}

For small rotation angle the offset momentum $\Delta K \rightarrow 0$ and perturbation theory in the dimensionless parameter
$\tilde t_{\perp}/\hbar v_F \Delta K$ fails. The breakdown of the perturbation theory occurs because of an incomplete treatment
of multiple scattering processes involving the interlayer coupling operator.  Recognizing this, Bistritzer and MacDonald (BM)
\cite{bismac1} developed a theory that treats multiple scattering through the three fundamental interlayer amplitudes that
describe a spatially modulated interlayer coupling with the period of the commensuration supercell. Their results show that the
Fermi velocity renormalization of the perturbation theory presages more dramatic physics at small rotation angle which can be
described as ``velocity reversal," i.e. the Fermi velocity changes {\it sign} as a function of (small) rotation angle crossing
through zero at a series of discrete magic angles. As a consequence the small angle regime is predicted to feature a manifold
of nearly flat bands at low energy.

The BM model is formulated as a two layer scattering theory: states with momentum $\vec k$ in one layer are scattered into
states at momentum $\vec k + \vec Q$ in its neighbor.  In the pseudospin basis the amplitudes for these processes are the $2
\times 2$ matrices given in Eqn. 18.  The gauge transformation $e^{- i \Delta \vec K \cdot \vec r}$ on the rotated layer brings
two Dirac nodes of neighboring layers into coincidence, and in this momentum shifted basis the three momentum transfers $\vec
Q_{i(=0,\pm 1)}$ are $Q_0=-\Delta \vec K$ and two $\pm 2 \pi/3$-rotated partners $Q_{\pm 1}$ which form a threefold symmetric
triad. Thus this construction considers a twofold layer-degenerate Dirac cone whose states are coupled through an off-diagonal
nonlocal operator containing three possible momentum transfers $\vec Q_i$ in the interlayer hopping. In the long distance
theory the single layer Hamiltonians are isotropic, so for {\it arbitrary} rotation angle $\theta$ the the theory is specified
by its unrenormalized Fermi velocity, the rotation angle and the coupling strength labelled $w$ in BM \cite{bismac1}, which
combine to form a single dimensionless scaling parameter $\alpha = w/2 \hbar v_F \sin(\theta/2)$.

Repeated action of the nonlocal interlayer operator generates a lattice of coupled momenta as shown in the top panel of Figure
6. The interlayer tunneling amplitudes are {\it directed transitions} in reciprocal space: the momenta $Q_i$ and their negates
$-Q_i$ describe complementary processes that transport electrons to and from the rotated layer. Consequently, an even number of
applications of the nonlocal operator to an initial single-layer Bloch state at wavevector $\vec k$ generates a Bloch state in
the same layer with momentum $\vec k + \vec {\cal Q}$ where $\vec {\cal Q}$ is spanned by the primitive vectors $\vec Q_1 -
\vec Q_0$ and $\vec Q_{-1} - \vec Q_0$. This defines a triangular reciprocal lattice whose six first star elements have
magnitude $\sqrt{3}\, Q_0$, and are $90^\circ$  rotated with respect to the original $\pm \vec Q_i$'s. An odd number of
applications of the operator transports the electron to the neighboring layer on the same reciprocal lattice, but offset by the
momentum shift $\vec Q_0$. The combination of these two sets describes a honeycomb lattice where the alternating sites
(momenta) occupy different layers as shown in Fig. 6.

\begin{figure}
\begin{center}
\includegraphics[angle=0,width=80mm,bb=0 0 600 800]{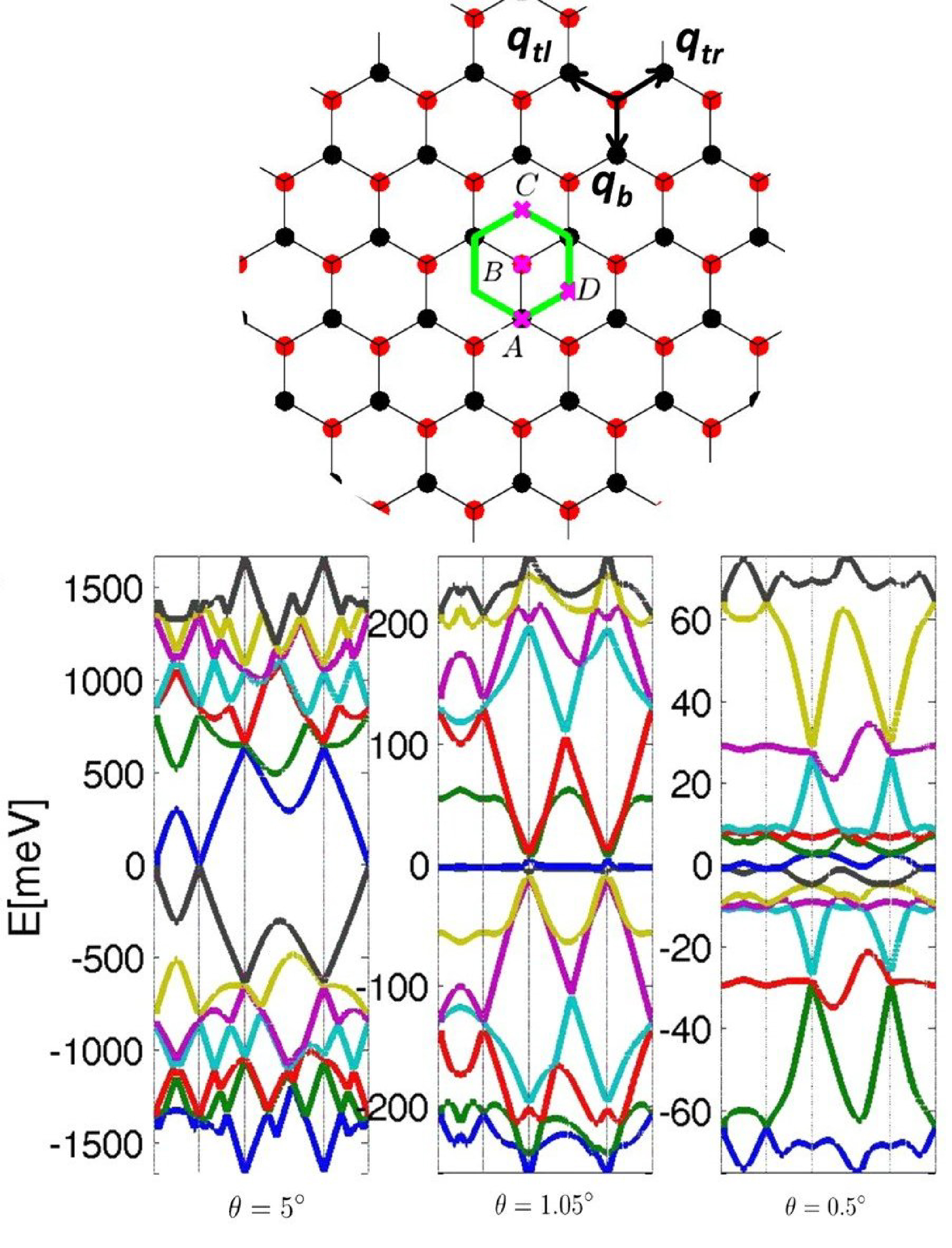}
\end{center}
\caption{\label{BMbands}  Top panel: A lattice of momenta is generated by repeated action of a nonlocal interlayer coupling
operator on a Bloch state in a single layer. The nonlocal operator transports an electron between layers and boosts the
momentum by a threefold symmetric triad of momentum transfers $Q_i$. An even number of applications of the operator generates a
triangular lattice of momenta in the original layer (red), an odd number generates a triangular lattice offset by momentum
$\Delta K$ (black). The combination forms a honeycomb lattice of coupled momenta. Bottom panel: Band structures obtained by
numerical diagonalization of the continuum Hamiltonian in a truncated plane wave basis retaining kinetic energies of order the
coupling strength $w$. The bands are plotted along the momentum space trajectory $ABCDA$ in the top figure. For small rotation
angles the bands flatten and the Fermi velocity of the zero energy states is strongly suppressed. Adapted from reference
\cite{bismac1}.}
\end{figure}

BM studied this model by numerically diagonalizing a truncated Hamiltonian expanded in a plane wave basis and retaining plane
waves with kinetic energies below the coherence scale $\sim w$. The effects of multiple scattering through the interlayer
coupling terms is then encoded in the structure of the bilayer eigenstates which contain coherent superpositions of the
single-layer Dirac modes. For large rotation angles ($\theta > 3^\circ$) the model reproduces the perturbation theory of Lopes
dos Santos et al. By contrast, in the very small angle regime the bandwidth $\hbar v_F Q_0$ collapses, the number of elements
in the low energy basis grows correspondingly and the electronic structure becomes spectrally congested as illustrated in Fig.
6. Thus the small angle regime is described by a strong coupled {\it multiband} theory that introduces physics inaccessible to
a low order perturbation theory. BM find that the low energy spectra in this regime show a very substantial reduction of the
Fermi velocity (typically $< 0.1$ of its single layer value) due to level repulsion among the coupled bands. Remarkably, the
reduced velocity parameter {\it oscillates} as a function of the fault angle as shown in Figure 7 and crosses zero at a series
of magic rotation angles. They suggest that this oscillation likely results from a $\theta$ dependence of the superpositions of
single layer modes that contain velocities of opposite sign, though a complete theory of the velocity oscillations has yet to
be developed. Thus the velocity renormalization found in the weak coupling limit represents just the first step towards a
complete twist-induced reconstruction of the low energy spectrum!

\begin{figure}
\begin{center}
\includegraphics[angle=0,width=80mm,bb=0 -50 1800 1200]{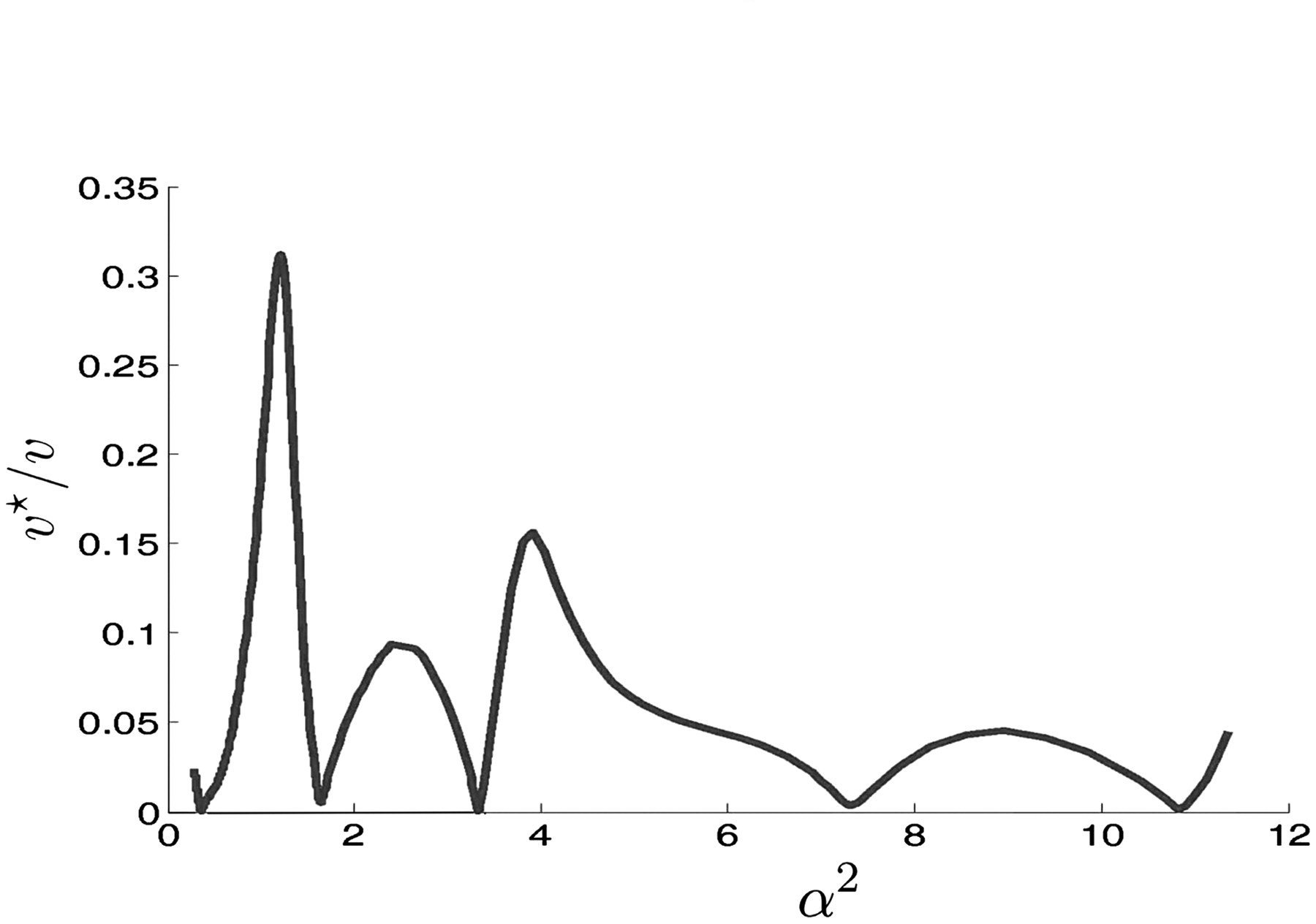}
\end{center}
\caption{\label{BMvel} The magnitude of the renormalized Fermi velocity for the zero energy states is plotted as a function of
the coupling parameter $\alpha^2 = (w/2 \hbar v_F \sin(\theta/2))^2$ for small rotation angles. The oscillations results from
sign changes of the Fermi velocity in the small rotation angle regime. For large rotation angles the renormalization factor
$v^*/v \approx 1 - 9 \alpha^2 \rightarrow 1$. Adapted from \cite{bismac1}.}
\end{figure}

\subsection{Two Band Models}

There has been interest in distilling the continuum theory to a simpler effective {\it two band} model that captures the
topological structure of its low energy spectrum. The approach is similar in spirit to the theory of Bernal stacked bilayers
\cite{McCFal,biasedbilayer,longbiasedbilayer} where one can integrate out its high energy degree of freedom to arrive at an
effective theory for its low energy states. For the Bernal bilayer this procedure identifies a new class of layer-coherent
chiral fermions that have quadratic low energy dispersion and a Berry's phase of $2 \pi$.  For the twisted bilayer, neglecting
commensuration effects, the low energy spectrum contains two layer-polarized {\it linear} Dirac cones that are recoupled at an
energy scale $\hbar v_F |\Delta \vec K|/2$ where they merge. The two band model attempts to provide a compact description of
the topological transition of the band dispersion that connects the low energy ``doubled cone" sector to its high energy
layer-hybridized sector.

For twisted bilayers the two band construction can be understood as a variant of the low energy theory for a Bernal bilayer
that allows for a finite momentum offset $|\Delta \vec K|$ between its Dirac nodes. Thus the low energy theory for Bernal
stacking is modified in the manner
\begin{eqnarray}
{\cal H}_K = -\frac{\hbar^2}{m} \left(%
\begin{array}{cc}
  0 & \partial^2 \\
  \bar \partial^2  & 0 \\
\end{array}%
\right) \rightarrow -\frac{\hbar^2}{m(\theta)} \left(%
\begin{array}{cc}
  0 & \partial^2 - (\Delta K)^2\\
  \bar \partial^2 - (\Delta \bar K)^2  & 0 \\
\end{array}%
\right)
\end{eqnarray}
valid for very small fault angles where $\hbar v_F |\Delta K|\ll \tilde t_\perp$ and $m=\tilde t_\perp/v_F^2$. This expression
can be derived by replacing the interlayer operators of Eqn. 18 by a simpler expression $\tilde t_\perp \sigma_-$ which
physically describes an interlayer tunneling amplitude across a single sublattice in each layer. The spectrum of this
Hamiltonian features a pair of Dirac cones, split by the momentum offset $|\Delta \vec K|$, that merge at a two dimensional
saddle point at $q=0$ representing the topological transition of the band structure. Importantly, the single layer Dirac cones
in this model have the {\it same} chirality so that annihilation of the Dirac points when they are coupled is topologically
forbidden. Generically, this model does allow for an energy offset between the Dirac points of the coupled bilayer but this is
believed to be small for physically reasonable coupling strengths.

The Hamiltonian in Eqn. 32 has been used to study the orbital quantization of a twisted bilayer in the presence of a
perpendicular magnetic field. By construction the limit $\Delta K=0$ describes the Landau quantization of Bernal bilayer
graphene: a Berry's phase of $2 \pi$ and quantized energy levels $\propto \sqrt{n(n-1)} B$. By contrast for finite rotation
angles the low energy spectrum of the offset model is a ``doubled" theory of single layer graphene: the fourfold degeneracy due
to the spin and valley degrees of freedom is doubled by an approximate layer decoupling of its low energy eigenstates. An
asymptotic analysis of the eigenvalues within this model demonstrates that splittings of the Landau level degeneracies due to
interlayer coupling are exponentially suppressed as a function of the rotation angle in the low energy regime \cite{saddleLL}.
The spectrum thus features a zero mode and Landau levels that disperse $\propto \sqrt{nB}$ \cite{saddledeGail,saddleLL}. This
twofold layer degeneracy is quickly eliminated as one passes through the crossover energy $\hbar v_F \Delta K/2$ where the
Dirac cones merge and hybridize. Above this crossover the spectrum has a different character: layer degeneracies are removed
and the quantized energies are $\propto (n+1/2)B$ as expected for a parabolic interlayer coherent band.

\section{Discussion}

Rotational faults commonly occur in a several different forms of graphene and their electronic properties are actively studied
experimentally. The rapidly growing experimental literature on this subject has not yet provided a unified picture of the
effects of faults on the electronic behavior, possibly due to differences in the electronic properties of samples produced by
different experimental methods.

A significant point of {\it agreement} among the various experimental works is that the interlayer coherence scale is very
small in these systems \cite{deheerreview,haas,graphenegraphite,miller}. This can be deduced clearly from their Landau level
spectra which have been measured by scanning tunneling spectroscopy (STS). These spectra show a scaling of the Landau level
energies $E_n \propto \sqrt{nB}$ \cite{graphenegraphite,miller} the signature of the Landau quantization of a massless Dirac
band, as observed for single layer graphene and quite distinct from the level sequence observed for Bernal stacked bilayers
\cite{McCFal}. Perhaps the strongest evidence for a reduction of the interlayer coupling scale comes from angle resolved
photoemission experiments which directly measure the quasiparticle dispersion relation \cite{sprinkle,hicks}  and find spectra
that follow the expected form for an isolated Dirac cone. These measurements have been interpreted as providing the first
direct measurement of the Dirac dispersion relation in graphene, uncontaminated by substrate or other interlayer effects
\cite{sprinkle}.

Since the effects of the interlayer coupling in twisted multilayers are intrinsically weak, their study is posing a significant
experimental challenge. It is here where different experiments carried out on different samples disagree. For example, the
Fermi velocity can be deduced from the slope of the $\sqrt{nB}$ scaling relation for the Landau quantized energies. The
strongest evidence for a twist-induced renormalization of $v_F$ comes from the Landau level spectra measured by scanning
tunneling spectroscopy of CVD graphenes grown on Ni substrates \cite{STMvel}. This work reports that $v_F$ is not constant as a
function of scanned position across a macroscopic sample, but instead it is found to vary in a range $0.87 \times 10^6 \, {\rm
m/s} < v_F < 1.1 \times 10^6 \, {\rm m/s}$. Simultaneous measurement of the topography of the Moire superlattice period of
these samples correlates the velocity reduction with the period and hence the rotation angle. The larger value, found for large
angle rotations, agrees well with the $v_F$ for single layer graphene and the $20 \%$ reduced value is correlated with a small
angle rotation $\sim 3^\circ$ as suggested by a perturbative analysis of the continuum theory \cite{JMLdS,bismac1}. This
contrasts with analogous STS measurements carried out for multilayer graphenes grown epitaxially on SiC ${\rm (000 \bar1)}$.
These also show the $\sqrt{nB}$ scaling of the Landau quantized energies. However, for these materials the slope of the scaling
relation yields a Fermi velocity $1.1 \times 10^6 \, {\rm m/s}$ for all samples studied down to a rotation angle of $1.4^\circ$
\cite{miller} completely spanning the range of rotation angles where a velocity renormalization is expected.

 A similar discrepancy arises in the spectroscopy of the van Hove singularity presumed to occur in the region where the momentum-offset Dirac cones of a twisted bilayer merge. Low
energy STS on Ni/CVD grown graphene reveals low energy peaks whose energies disperse with their topographically measured
rotation angles in the low angle regime $1.2 < \theta < 3.5^\circ$ \cite{vanHove} roughly consistent with the van Hove
scenario. Yet these features are not seen at all in spectroscopy of the SiC  epitaxial twisted graphenes regardless of the
fault angle. Perhaps the strongest challenge to the idea of a twist-induced spectral reconstruction comes from angle resolved
photoemission. These measurements directly measure the quasiparticle dispersion and clearly resolve the Dirac cone with a Fermi
velocity that is indistinguishable from that of single layer graphene. Despite a careful search, no evidence is found in these
measurements for any type of hybridization between Dirac cones in the spectral regions where they cross \cite{hicks}. The
simplest interpretation of the ARPES data is that the first few graphene layers accessible to this spectroscopy are
electronically floating, i.e. extremely weakly coupled to each other and to deeper layers in the film.

An important goal for theory in this area is therefore to identify situations where the effects of the interlayer coupling
across a rotational fault are manifested in their electronic behavior. There has been progress in this direction. Bistritzer
and MacDonald have studied the effect of the rotation angle of a bilayer on its interlayer tunneling conductance, predicting
dramatic enhancements of the vertical conductance at special rotation angles that can be identified with low order commensurate
superlattices \cite{rafibilayer}. More recent work has pointed to nontrivial effects of a $\theta$-dependent interlayer
coupling on the equilibrium charge redistribution across the bilayer in a perpendicular field \cite{brey}.  Kindermann and I
studied the Landau level spectra for weakly coupled bilayers and find that even a weak coherence splitting of bilayer bands at
energies well above the mass scale produces a striking new effect, the Dirac comb \cite{mkejm}. Here small differences in the
orbitally quantized states in two weakly coherence-split bands produces an amplitude modulation of the Landau level spectrum
with a period that greatly exceeds the coherence scale, and should be observable by magnetotransport in the weak field regime.
Small rotations angles can introduce long spatial Moire periods for twisted bilayers that can be made commensurate with the
magnetic length $\sqrt{\hbar/eB}$ on accessible field scales, accessing Hofstadter commensuration physics in a new family of
materials \cite{rafibutterflies}. The band flattening theoretically predicted for in the small twist angle regime will surely
focus attention on many body effects in the low energy physics. Further studies along all of these lines provide a very open
area for further work.

One might be discouraged by the lack of a definitive theory of the electronic structure of rotationally faulted graphenes. To
the contrary this is an exciting situation. These systems are challenging to the most familiar tools of electronic structure
theory and their understanding is likely to involve creative new approaches.

\section*{Acknowledgements}

My research that is reported in this review is supported by the Department of Energy under contract DE-FG02-ER45118.

\end{document}